\documentclass[aps,prx,twocolumn,superscriptaddress,showpacs,longbibliography]{revtex4-1}
\usepackage{amsmath}
\usepackage{graphicx}
\usepackage{amsfonts}
\usepackage{verbatim}
\usepackage{amssymb}
\usepackage{color}
\usepackage{dsfont}
\usepackage{amsmath} 
\usepackage{hyperref}
\usepackage{physics}
\hypersetup{colorlinks = true, urlcolor = blue, linkcolor = blue, citecolor = blue}

\newcommand{\su}{\uparrow} 
\newcommand{\sd}{\downarrow} 
\newcommand{\bpm}{\begin{pmatrix}}
\newcommand{\epm}{\end{pmatrix}}

\newcommand{\nn}{\nonumber \\} 

\newcommand{\dg}{^{\dagger}}

\begin{document}

\title{Scaling up a sign-ordered Kitaev chain without magnetic flux control}

\author{Chun-Xiao Liu}\email{Corresponding author: chunxiaoliu62@gmail.com}
\affiliation{QuTech and Kavli Institute of Nanoscience, Delft University of Technology, Delft 2600 GA, The Netherlands}

\author{Sebastian Miles}
\affiliation{QuTech and Kavli Institute of Nanoscience, Delft University of Technology, Delft 2600 GA, The Netherlands}

\author{Alberto Bordin}
\affiliation{QuTech and Kavli Institute of Nanoscience, Delft University of Technology, Delft 2600 GA, The Netherlands}

\author{Sebastiaan L. D. ten Haaf}
\affiliation{QuTech and Kavli Institute of Nanoscience, Delft University of Technology, Delft 2600 GA, The Netherlands}

\author{Grzegorz P. Mazur}
\affiliation{QuTech and Kavli Institute of Nanoscience, Delft University of Technology, Delft 2600 GA, The Netherlands}

\author{A. Mert Bozkurt}
\affiliation{QuTech and Kavli Institute of Nanoscience, Delft University of Technology, Delft 2600 GA, The Netherlands}

\author{Michael Wimmer}
\affiliation{QuTech and Kavli Institute of Nanoscience, Delft University of Technology, Delft 2600 GA, The Netherlands}

\date{\today}

\begin{abstract}
Quantum dot-superconductor arrays have emerged as a new and promising material platform for realizing topological Kitaev chains.
So far, experiments have implemented a two-site chain with limited protection.
Here we propose an experimentally feasible protocol for scaling up the chain in order to enhance the protection of the Majorana zero modes.
To this end, we make use of the fact that the relative sign of normal and superconducting hoppings mediated by an Andreev bound state can be changed by electrostatic gates.
In this way, our method only relies on the use of individual electrostatic gates on hybrid regions, quantum dots, and tunnel barriers, respectively, without the need for individual magnetic flux control, greatly simplifying the device design.
Our work provides guidance for realizing a topologically protected Kitaev chain, which is the building block of error-resilient topological quantum computation.
\end{abstract}

\maketitle

\emph{Introduction.}---The Kitaev chain is a paradigm of topological superconductivity that can host Majorana zero modes~\cite{Kitaev2001Unpaired,Alicea2012New, Leijnse2012Introduction, Beenakker2013Search, Stanescu2013Majorana, Jiang2013Non, Elliott2015Colloquium, Sato2016Majorana, Sato2017Topological, Aguado2017Majorana, Lutchyn2018Majorana, Zhang2019Next, Prada2020From, Frolov2020Topological, Flensberg2021Engineered}.
These zero-energy excitations are non-Abelian anyons which can be utilized to implement topological quantum computation~\cite{Nayak2008Non-Abelian, DasSarma2015Majorana}. 
Recently, quantum dot-superconductor arrays have emerged as a promising platform for realizing  a Kitaev chain~\cite{Sau2012Realizing}.
A minimal two-site version~\cite{Leijnse2012Parity} has been successfully realized in low-dimensional semiconductors, supported by tunnel spectroscopic evidence of Majorana zero modes at a fine-tuned sweet spot~\cite{Dvir2023Realization, tenHaaf2024Twosite, Zatelli2024Robust}.
Crucially, a balance of the normal and superconducting coupling strengths is achieved by electrostatic gating on the hybrid region~\cite{Liu2022Tunable, Bordin2023Tunable}.
However, these finely-tuned zero modes remain vulnerable to environmental noises due to a limited protection~\cite{Leijnse2012Parity, Liu2022Tunable, Tsintzis2022Creating, Liu2024Enhancing, Souto2023Probing, Miles2024Kitaev, Luna2024Flux, Samuelson2024Minimal, Souto2024Subgap}, which can be enhanced and become topological only after the quantum dot array is scaled up~\cite{Kitaev2001Unpaired, Sau2012Realizing, Ezawa2024Even_odd, Bordin2024Crossed, Bordin2024Signatures, tenHaaf2024Edge}.
Furthermore, the experiments of anyonic fusion and braiding~\cite{Liu2023Fusion, Pandey2024Nontrivial, Boross2024Braiding, Tsintzis2024Majorana} to detect the non-Abelian statistics would not be possible before the Kitaev chain is extended to four or more sites.

\begin{figure}
\centering
{\includegraphics[width = \linewidth]{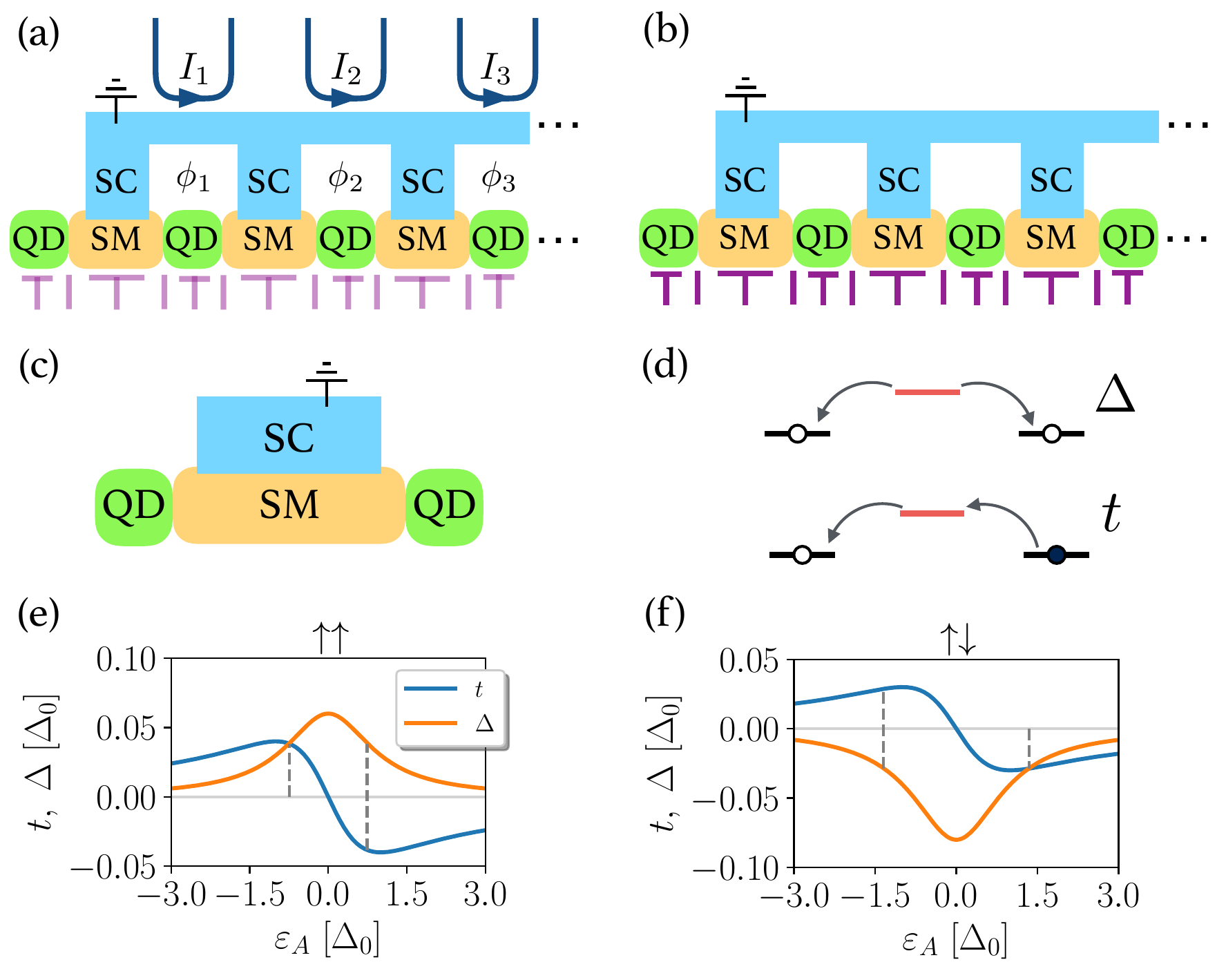}}
\caption{
(a) Schematic of a device with magnetic flux control using bias lines.
(b) Schematic of a device where the phase difference is controlled purely by electrostatic gates (purple lines).
(c) and (d) Schematic of a two-site Kitaev chain and the virtual processes that induce effective couplings of $\Delta$ and $t$.
(e) and (f) Dependence of the coupling amplitudes on the chemical potential of the Andreev bound states.}
\label{fig:Fig_1}
\end{figure}

In an extended chain ($N \geq 3$), the phases of the couplings become particularly important.
In the limit of confinement to a one-dimensional channel as in experiments \cite{Dvir2023Realization, tenHaaf2024Twosite, Zatelli2024Robust, Wang2022Singlet, Wang2023Triplet} and in the presence of Rashba spin-orbit interaction and an axial magnetic field, an approximate complex conjugate symmetry~\cite{Tewari2012Topological} further constrains the effective couplings to be real numbers~\cite{Sau2012Realizing, Scheid2009Anisotropic, Diez2012Andreev}. 
Thus, the problem of phase uncertainty is now reduced to \emph{sign} uncertainty.
For an $N$-site Kitaev chain
\begin{align}
H_{K} = \sum^N_{n=1} \varepsilon_n f\dg_n f_n + \sum^{N-1}_{n=1} (t_n f\dg_{n+1} f_n + \Delta_n f\dg_{n+1} f\dg_n + h.c.),
\label{eq:spinless_kitaev}
\end{align}
the sweet spot condition becomes
\begin{align}
& \varepsilon_n=0,\quad \abs{t_n}=\abs{\Delta_n},\quad \text{sgn}(t_1 \Delta_1) = \text{sgn}(t_n \Delta_n),
\end{align}
where $f_n$ is the annihilation operator of a spinless fermion, $\varepsilon_n$ is the onsite energy, and $t_n$ and $\Delta_n$ are the amplitudes of normal and superconducting tunnelings, respectively.
In Refs.~\cite{Sau2012Realizing}, the proposed solution to the sign problem was to use an individual magnetic flux control of the phase between neighboring superconducting grains.
However, this would inevitably introduce multiple flux bias lines, thus complicating the device design and causing heating problems~[see Fig.~\ref{fig:Fig_1}(a)].
In particular, cross-talk of flux bias lines becomes an issue when using small superconducting loops, while larger-size loops would significantly increase the device size and thus limit the possible number of quantum dots to scale up.

In the current work, we propose a scale-up protocol, where the sign problem is fixed purely in an electrostatic way \emph{without} magnetic flux control~[see Fig.~\ref{fig:Fig_1}(b)].
The new physical insight here is that the two sweet spots mediated by an Andreev bound state (ABS) have opposite signs and can be explicitly detected in a three-site setup by conductance spectroscopy.
Since a set of electrostatic gates is always needed to individually control $\varepsilon_n, \abs{t_n}$ and $\abs{\Delta_n}$ in a Kitaev chain, our proposal does not introduce any additional overhead in the device fabrication.
Instead our method greatly simplifies the device design and makes the platform suitable for implementing scalable topological quantum computation.

\emph{Sign of sweet spot.}---We first consider a minimal setup consisting of double quantum dots connected by a hybrid segment.
The Hamiltonian is~\cite{Dominguez2016Quantum, Liu2022Tunable, Bordin2023Tunable, Tsintzis2022Creating, Liu2024Enhancing}
\begin{align}
& H = H_{D,1} + H_{A,1} + H_{D,2} + H_{DAD, 1}, \nn
& H_{D,i} = \sum_{\sigma =\su, \sd}(\varepsilon_{Di} + \sigma E_{Zi}) n_{Di\sigma} + U_{Di} n_{Di\su}n_{Di\sd}, \nn
& H_{A,i} = \sum_{\sigma =\su, \sd} \varepsilon_{Ai} n_{Ai\sigma} + ( \Delta_{ i} c_{Ai\su}c_{Ai\sd} + h.c.), \nn
& H_{DAD,i} =  \sum_{\sigma=\su,\sd} \Big(
t_{\text{sc}, i} c\dg_{Ai\sigma} c_{Di\sigma} + t'_{\text{sc}, i}  c\dg_{Di+1\sigma} c_{Ai\sigma} \nn
&\quad + \sigma t_{\text{sf}, i} c\dg_{Ai\overline{\sigma}} c_{Di\sigma} + \sigma t'_{ \text{sf}, i} c\dg_{Di+1\overline{\sigma}} c_{Ai\sigma} + h.c. \Big).
\label{eq:H_K2}
\end{align}
Here $H_D$ is the Hamiltonian for a quantum dot, $\varepsilon_{D}$ is the orbital energy, $E_{Z}$ is the induced Zeeman spin splitting, and $U_D$ is the Coulomb repulsion. 
$H_{A}$ is the Hamiltonian of a subgap ABS in the hybrid region, $\varepsilon_{A}$ is the normal-state energy and $\Delta$ is the induced pairing.
$H_{DAD}$ describes single electron tunneling between dots and hybrids, $t_{\text{sc}}$ ($t_{\text{sf}}$) is the amplitude for spin-conserving (spin-flipping) processes.
When the direction of the spin-orbit field is perpendicular to the applied magnetic field~\cite{Wang2022Singlet, Wang2023Triplet}, $t_{\text{sc}}, t_{\text{sf}}$ are real~\cite{Tewari2012Topological,Sau2012Realizing}.
Here, we assume a single dot orbital and single ABS in $H_D$ and $H_A$, respectively. 
This approximation is accurate when the level spacings are large, i.e., $\Delta \varepsilon_D > E_Z$ and $\Delta \varepsilon_{A} > t_{\rm{sc}}, t_{\rm{sf}}$, as discussed in Ref.~\cite{Liu2022Tunable} and demonstrated in recent experiments~\cite{Bordin2023Tunable, Zatelli2024Robust}.

In the tunneling regime where $\abs{t_{\text{sc}}}, \abs{t_{\text{sf}}} \ll \Delta, E_Z$, the effective couplings can be obtained using perturbation theory 
\begin{align}
& t_{\su \su} = (   t_{\rm{sf},1}  t'_{\rm{sf},1} - t_{\rm{sc},1}  t'_{\rm{sc},1} ) \frac{u^2 - v^2}{E_A}, \nn
& \Delta_{\su \su} = ( t_{\rm{sc},1} t'_{\rm{sf},1} + t_{\rm{sf},1} t'_{\rm{sc},1} )  \frac{2uv}{E_A},
\label{eq:eff_coupling_upup}
\end{align}
where $t_{\su \su}$ and $\Delta_{\su \su}$ are the effective normal and superconducting couplings between spin-up orbitals in two quantum dots.
$u^2=1-v^2 =1/2 + \varepsilon_{A}/2E_A$ are the coherence factors, and $E_A=\sqrt{\varepsilon^2_A+\abs{\Delta_0}^2}$ is the excitation energy.
Figure~\ref{fig:Fig_1}(e) shows the dependence of $t_{\su \su}$ and $\Delta_{\su \su}$ on the chemical potential of the hybrid region, with model parameters $\Delta_1=\Delta_0$, $t_{\rm{sc}}=3t_{\rm{sf}}=0.3\Delta_0$.
Here both amplitudes are real due to complex conjugate symmetry~\cite{Tewari2012Topological}, and furthermore, the two sweet spots have opposites signs, i.e.,
\begin{align}
& t_{\su \su} = \Delta_{\su \su},~\text{for}~\varepsilon_A = - \varepsilon^*_A, \nn
& t_{\su \su} = -\Delta_{\su \su},~\text{for}~\varepsilon_A = \varepsilon^*_A,
\label{eq:t_up_up}
\end{align}
where $\varepsilon^*_A=\Delta_0 (t_{\rm{sc},1} t'_{\rm{sf},1} + t_{\rm{sf},1} t'_{\rm{sc},1})/(t_{\rm{sf},1}  t'_{\rm{sf},1} - t_{\rm{sc},1}  t'_{\rm{sc},1})=0.75\Delta_0$.
We emphasize that the existence of two opposite-sign sweet spots is a robust feature as evidenced in Eq.~\eqref{eq:eff_coupling_upup}.
For example, when the strength of spin-orbit interaction becomes much stronger ($3t_{\rm{sc}} = t_{\rm{sf}}=0.3\Delta_0$), the only effect is that $t_{\su\su}$ obtains an overall minus sign, thus only reversing the signs of the sweet spots relative to Fig.~\ref{fig:Fig_1}(e).
In addition, change of the parity of the bound-state wavefunctions ($t \to -t$ or $t' \to -t'$) would only give a common minus sign to both $t_{\su \su}$ and $\Delta_{\su \su}$, not affecting the sweet spot properties either.
On the other hand, the coupling amplitudes between orbitals of opposite spins are
\begin{align}
&t_{\su \sd} = -(  t_{\rm{sc},1} t'_{\rm{sf},1} + t_{\rm{sf},1} t'_{\rm{sc},1} ) \frac{u^2 -v^2}{E_A}, \nn
&\Delta_{\su \sd} =  ( t_{\rm{sf},1} t'_{\rm{sf},1}-  t_{\rm{sc},1} t'_{\rm{sc},1}  ) \frac{2uv}{E_A}.
\label{eq:t_up_down}
\end{align}
Figure~\ref{fig:Fig_1}(f) shows the $t_{\su \sd}$ and $\Delta_{\su \sd}$ curves using the same model parameters as Fig.~\ref{fig:Fig_1}(e).
Now two sweet spots appear at $\varepsilon_A = \pm \varepsilon^*_A$ with $\varepsilon^*_A= \Delta_0 ( t_{\rm{sf},1} t'_{\rm{sf},1}-  t_{\rm{sc},1} t'_{\rm{sc},1}  )/(  t_{\rm{sc},1} t'_{\rm{sf},1} + t_{\rm{sf},1} t'_{\rm{sc},1} ) = 4\Delta_0/3$, and interestingly, their signs are reversed relative to the same-spin scenario.
Here in obtaining Eqs.~\eqref{eq:t_up_up} and~\eqref{eq:t_up_down}, we have assumed no Zeeman splitting in ABS, which in general is expected to be reduced due to $g$ factor renormalization~\cite{Stanescu2010Proximity,Antipov2018Effects}.
Nevertheless, even when this assumption is relaxed, it is still possible to find two opposite-sign sweet spots due to continuity~\cite{Liu2022Tunable, supp_scaling}.
Hence, we find that generically the relative sign of the normal and superconducting couplings can be changed by either changing the chemical potential in the hybrid region to switch to the other sweet spot or by changing the dot energy to switch the spin polarization, with both ways using electrostatic gating only.

\begin{figure*}
\centering
{\includegraphics[width = 1.0 \textwidth]{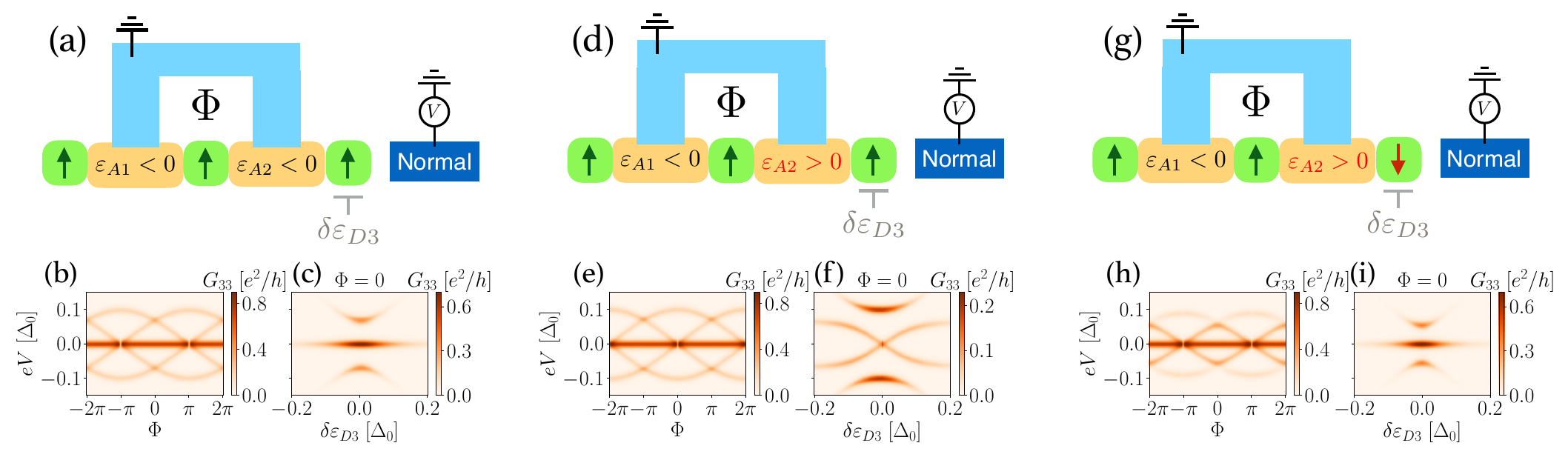}}
\caption{Upper panels: Schematics of the transport setup to detect the signs of the sweet spots.
Lower panels: Conductance spectroscopy ($G_{33}=dI_3/dV_3$) as a function of the magnetic flux or dot detuning.
The systems in panels (a) and (g) are sign-ordered Kitaev chains at $\Phi=0$, while in panel (d) the two sweet spots have opposite signs.}
\label{fig:Fig_2}
\end{figure*}

\emph{Detection of $\pi$-phase shift.}---To experimentally detect the subtle sign of sweet spots, the minimal setup is a three-quantum-dot device with a superconducting loop connecting the two hybrid regions [see Fig.~\ref{fig:Fig_2}(a)].
Tunnel spectroscopy would distinguish the signs of the sweet spots by a $\pi$-phase shift.
To support the statement, we now perform numerical calculations using the following Hamiltonian
\begin{align}
H =& H_{D1} + H_{A1} + H_{D2} + H_{A2} \nn
&+ H_{D3} + H_{DAD,1} + H_{DAD,2} 
\label{eq:H_K3}
\end{align} 
which includes three normal quantum dots connected by two ABSs.
The Hamiltonians for dots, ABSs and electron tunneling are almost identical to those in Eq.~\eqref{eq:H_K2}, except that now for $H_A$ a phase difference determined by the magnetic flux is included in the pairing potential, i.e., $\Delta_1 = \Delta_0,~\Delta_2 = \Delta_0e^{i\Phi}$.
In addition, a normal-metal lead is tunnel coupled to dot $D3$, and conductance is numerically calculated using the rate-equation method~\cite{Beenakker1991Theory, Bruus2004Many}.

We first consider a scenario where the two sweet spots are of the same type.
By setting all three quantum dots to be spin-up, this condition is satisfied when $\varepsilon_{A1} = \varepsilon_{A2} \approx -0.804\Delta_0$, which is close to the values predicted in Eq.~\eqref{eq:t_up_up}.
Figure~\ref{fig:Fig_2}(b) shows that a sign-ordered three-site Kitaev chain indeed appears at $\Phi=0$ with a stable and isolated zero-bias conductance peak induced by Majorana zero modes.
Additionally, this zero-bias peak is robust against detuning of dot $D3$, see Fig.~\ref{fig:Fig_2}(c), verifying that Majoranas are spatially localized.

When we change $\varepsilon_{A2}\approx 0.704\Delta_0$ while keeping $\varepsilon_{A1}$ unchanged, the sign of the sweet spot mediated by $A2$ becomes opposite to $A1$, see Fig.~\ref{fig:Fig_1}(e).
As shown in Fig.~\ref{fig:Fig_2}(e), an additional zero-energy state appears in the vicinity of $\Phi=0$, making the system gapless~\cite{Pandey2024Nontrivial}.
Unlike the sign-ordered chain, now the zero-bias peak is readily split with detuning of $D3$, see Fig.~\ref{fig:Fig_2}(f), due to the hybridization between the Majorana and the additional zero-energy state~\cite{supp_scaling}.
Moreover, by comparing Figs.~\ref{fig:Fig_2}(b) and~\ref{fig:Fig_2}(e) the sign switch of sweet spot is clearly revealed as a $\pi$-phase shift in the flux-dependent conductance spectroscopy.

In the third scenario, we flip the spin of $D3$ into spin-down, which can be experimentally implemented by electrostatic gating.
The chemical potential of $A2$ is still positive: $\varepsilon_{A2} \approx 1.3\Delta_0$.
Indeed, Figs.~\ref{fig:Fig_2}(h) and~\ref{fig:Fig_2}(i) shows the emergence of a sign-ordered Kitaev chain again, confirming the predictions made in Fig.~\ref{fig:Fig_1}(f).
Thereby flipping the spin of the dot orbitals provides an additional knob for correcting the sign of sweet spots.

\begin{figure*}
\centering
{\includegraphics[width =0.8\textwidth]{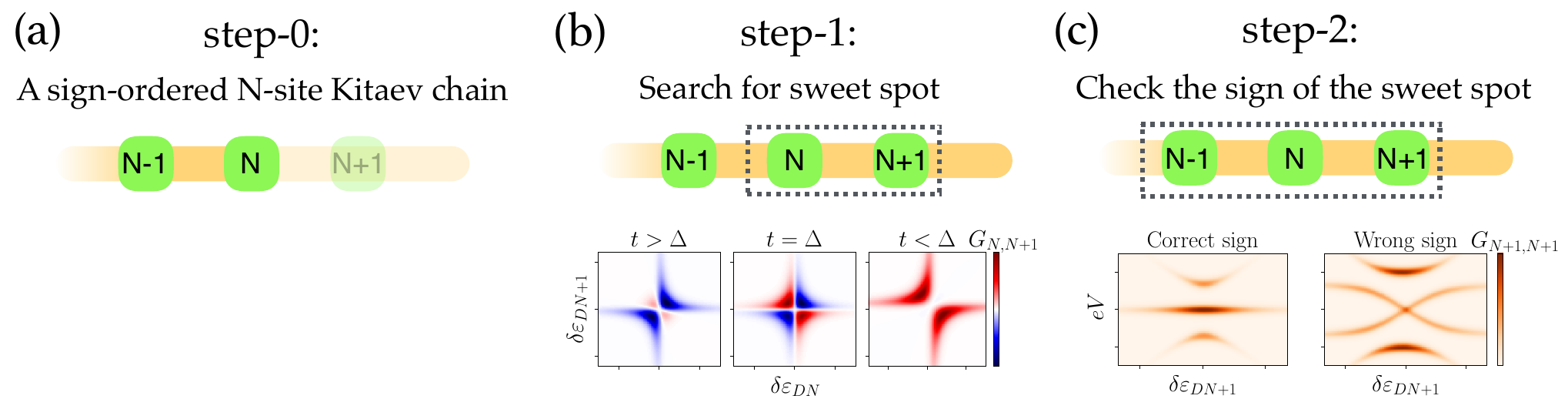}}
\caption{
Protocol for tuning up a sign-ordered Kitaev chain.
(a) Preparation: get ready a sign-ordered $N$-site chain ($N \geq 2$).
(b) Step-1: Switch on the coupling between the $N$ and $N+1$-th dots while decoupling them from the rest of the system, e.g., by closing the tunnel barriers indicated by the vertical lines of the rectangle or by detuning all the other dot orbitals off resonance.
Find the sweet spot $|t|=|\Delta|$ in the charge stability diagram.
(c) Step-2: Connect the $N+1$-th dot with the $N$- and $N-1$-th dots, and measure the differential conductance against the detuning of dot-$N+1$.
If the zero-bias conductance peak remains robust, we thereby obtain a sign-ordered $N+1$-site Kitaev chain.
Otherwise, we should return to step-1 to find a new sweet spot and test it in step-2 until success.
}
\label{fig:Fig_3}
\end{figure*}

\emph{Protocol for scaling up.}---Based on the findings in the previous sections, we now put forward our protocol for scaling up a long sign-ordered Kitaev chain.
To this end, we require an experimental setup that can (i) be used to tune two neighboring dots to a sweet spot, for example as discussed in~\cite{Dvir2023Realization,Zatelli2024Robust,tenHaaf2024Twosite,Liu2024Enhancing}, and (ii) detect whether the zero-energy degeneracy splits when the energy of the final dot is detuned from the resonance.
In general, this requires that the superconducting leads that proximitize different hybrid regions form a single grounded lead. 
The two measurements can be realized for example by coupling each normal dot to an individual normal lead, forming a multi-terminal junction.
Alternatively, it is also sufficient to only contact the final dot with a normal lead, as shown in Fig.~\ref{fig:Fig_2} or using gate sensing.
Our protocol allows to build up the chain iteratively dot by dot.

\emph{Step-0:} To begin with, we assume that we have already obtained a sign-ordered $N$-site Kitaev chain ($N\geq 2$) as shown in Fig.~\ref{fig:Fig_3}(a) (for $N=2$, this corresponds to finding the sweet spot).
Our goal is to extend it to $N+1$ sites by choosing an appropriate sweet spot for the newly added dot.

\emph{Step-1:} First, we focus on a two-site system formed by the $N$- and $N+1$-th quantum dots decoupled from the rest of the array [see the dashed rectangle in Fig.~\ref{fig:Fig_3}(b)].
This can be achieved by closing the tunnel barriers outside the two dots, or, alternatively, by shifting all the other dots off-resonance, as illustrated in the experiments of Refs.~\cite{Bordin2024Signatures, tenHaaf2024Edge}.
Then by electrostatic gating on the hybrid region, a sweet spot with $|t_N| = |\Delta_N|$ can be reached, e.g. signified by a cross in the charge stability diagram, see Fig.~\ref{fig:Fig_3}(b)~\cite{Dvir2023Realization,Zatelli2024Robust,tenHaaf2024Twosite}.
However, the sign of the sweet spot remains uncertain so far.

\emph{Step-2:} We then form an three-site chain by coupling the $N-1$-, $N$-, and $N+1$-th dots, e.g. by lowering tunnel barriers or by bringing the $N-1$-th dot back to resonance, see Fig.~\ref{fig:Fig_3}(c).
We measure the conductance spectroscopy against the detuning of dot-$N+1$, see Fig.~\ref{fig:Fig_3}(c).
If the zero-bias peak is robust, we have successfully extended an $N$-site chain to $N+1$ and can continue with the next dot.
Otherwise, we have to return to step-1 to tune to the other sweet spot or the other dot spin, effectively flipping the relative sign between $t_N$ and $\Delta_N$.

\begin{figure}[t]
\centering
{\includegraphics[width = \linewidth]{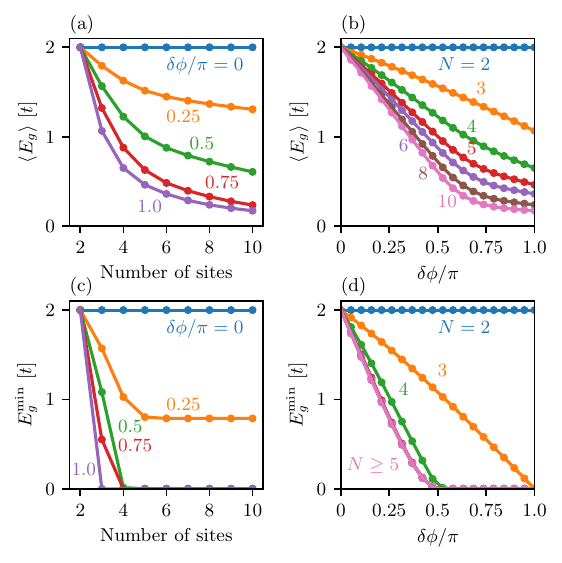}}
\caption{
(a) and (b) Mean of excitation gap of a Kitaev chain versus number of sites and phase fluctuation amplitude.
(c) and (d) Minimal excitation gap versus number of sites and phase fluctuation amplitude.
}
\label{fig:Fig_4}
\end{figure}

\emph{Effect of phase fluctuations}---In a realistic device, complex conjugate symmetry can be broken due to a finite width of the one-dimensional channel, the magnetic orbital effect on the dot or ABS wavefunctions, or a misaligned magnetic field.
After performing a gauge transformation, an $N$-site chain ($N \geq 3$) can have $N-2$ independent phase fluctuations i.e. $\Delta_i = \abs{\Delta_i}e^{i\delta \phi_i}$ for $i=2,3,\ldots, N-1$~\cite{supp_scaling}.
Here we focus on the energy gap in the presence of phase fluctuations, since $\abs{t_i}=\abs{\Delta_i}$ guarantees the presence of a zero energy.
Here phase fluctuation obeys a uniform distribution of $\delta \phi_i \in [-\delta \phi, \delta \phi]$ for an ensemble of size 2000.
As shown in Figs.~\ref{fig:Fig_4}, both the averaged and the minimal gap decreases monotonically with either the number of sites or the phase fluctuation amplitude.
Interestingly, a longer chain is more prone to becoming gapless when $\delta \phi > \pi/2$.
We hence expect that our protocol remains applicable even for small deviations from one dimensionality, as the sign of the sweet spot can still be used to minimize the phase difference.
Put in another way, reducing the cross-section area of the semiconducting nanowires would suppress the phase deviations induced by the magnetic orbital effects~\cite{Antipov2018Effects}, thus optimizing the protection of Majorana zero modes.

\emph{Discussion.}---In previous works, it was suggested to satisfy the phase match condition in an extended Kitaev chain by controlling the magnetic flux through a superconducting loop~\cite{Sau2012Realizing,Fulga2013Adaptive}.
However, devices with small loops [see Fig.~\ref{fig:Fig_1}(a)] would suffer from cross-talk issues of the flux bias lines, while larger loops would significantly increase the system size by tens of micrometers, making it difficult to fit into a nanoscale device and limiting the number of dots to scale up.
Within this context, our scale-up proposal uses a purely electrostatic method, thus eliminating the need of the cumbersome magnetic flux control.
Most crucially, such a gate configuration is needed anyways to fine-tune the sweet-spot values of $\varepsilon_n=0$ and $| t_n |= | \Delta_n |$, so it doesn't add any fabrication overhead to the device design.
Practically, the electrostatic gate configurations have already been implemented as a set of finger gates in recent experiments of three-site chains~\cite{Bordin2024Signatures, tenHaaf2024Edge}, which can be further generalized to longer chains.
Thereby, we note the crucial difference between the upscaling protocol proposed in Ref.~\cite{Fulga2013Adaptive} and ours is that here we rely only on the electrostatic gates that are already present in the device. 
There is no need of additional gates or additional current lines for magnetic flux control.


\emph{Summary.}---In this work, we discover that the sweet spot mediated by ABS has a \emph{sign} uncertainty, which was overlooked and undetectable in two-site chain studies~\cite{Liu2022Tunable, Tsintzis2022Creating, Liu2024Enhancing, Souto2023Probing, Dvir2023Realization, tenHaaf2024Twosite, Zatelli2024Robust, Liu2024Coupling, Bozkurt2024Interaction, vanDriel2024Cross}, but will become crucial in an extended chain.
Based on that, we give a concrete and practical protocol for scaling up a Kitaev chain using only electrostatic gates, eliminating the need of a magnetic flux control.
It avoids the adverse heating issue and at the same time maintains a small nanoscale device size, both of which will benefit the eventual implementation of a scalable topological quantum computer. 
In particular, the gate configuration and control required in our proposal have been implemented in a recent experiment~\cite{tenHaaf2024Edge}, adding to the practicality of our work.
We thus believe that our proposal provides guidance to realizing a long topological Kitaev chain for implementing topological quantum computing~\cite{Nayak2008Non-Abelian, DasSarma2015Majorana} as well as demonstrating the non-Abelian statistics of Majorana anyons~\cite{Liu2023Fusion, Boross2024Braiding, Tsintzis2024Majorana}.

\begin{acknowledgements}
We are grateful to N. van Loo, F. Zatelli, and L. P. Kouwenhoven for useful discussions. 
This work was supported by a subsidy for top consortia for knowledge and innovation (TKI toeslag), by the Dutch Organization for Scientific Research (NWO) through OCENW.GROOT.2019.004, and by Microsoft Corporation Station Q.

\emph{Author contributions.}---
C.-X.L. conceived the project idea and designed the project with input from A.B., S.L.D.t.H., G.P.M., and M.W.
C.-X.L. carried out the calculations with input from S.M. and A.M.B.
C.-X.L. and M.W. wrote the manuscript with input from all authors.
C.-X.L. and M.W. supervised the project.
\end{acknowledgements}

\bibliography{references_CXL}

\appendix
\onecolumngrid
\vspace{1cm}
\begin{center}
{\bf\large Supplemental Materials for ``Scaling up a sign-ordered Kitaev chain without magnetic flux control''}
\end{center}
\vspace{0.5cm}

\setcounter{secnumdepth}{3}
\setcounter{equation}{0}
\setcounter{figure}{0}
\renewcommand{\theequation}{S-\arabic{equation}}
\renewcommand{\thefigure}{S-\arabic{figure}}
\renewcommand\figurename{Supplementary Figure}
\renewcommand\tablename{Supplementary Table}
\newcommand\Scite[1]{[S\citealp{#1}]}
\newcommand\Scit[1]{S\citealp{#1}}
\makeatletter \renewcommand\@biblabel[1]{[S#1]} \makeatother


\section{Model parameter values for numerical calculations}

The Hamiltonian parameters we choose for numerical calculations of tunnel conductances are (in unit of the induced gap $\Delta_0=1$): 
\begin{align}
& |\Delta_i| = \Delta_0 = 1, \quad  t_{\text{sc},i}=0.3, \quad t_{\text{sf},i} = 0.1, \quad E_{Zi}=2, \nn
& U_{Di}=5,\quad \Gamma=0.005, \quad k_BT=0.005,
\end{align}
where $\Gamma$ is the dot-lead coupling strength.
The parameters are chosen corresponding to the experimental devices recently studied in Refs.~\cite{Wang2022Singlet, Wang2023Triplet, Bordin2023Tunable, Dvir2023Realization, tenHaaf2024Twosite, Zatelli2024Robust, Bordin2024Crossed,Bordin2024Signatures}.

\section{Gauge transformation on Kitaev chain}\label{sec:gauge}
The Hamiltonian of an $N$-site spinless Kitaev chain is given by
\begin{align}
H_{K} = \sum^N_{n=1} \varepsilon_n f\dg_n f_n + \sum^{N-1}_{n=1} (t_n e^{i\alpha_n} f\dg_{n+1} f_n + \Delta_n e^{i\beta_n} f\dg_{n+1} f\dg_n + h.c.).
\end{align}
Here $\varepsilon_n, t_n, \Delta_n >0$ and $0 \leq \alpha_n, \beta_n < 2\pi$ without loss of generality. 
We now perform a gauge transformation on the spinless fermion operators as below
\begin{align}
& \widetilde{f}_n = f_n e^{i\theta_n}, \quad  \widetilde{f}\dg_n = f\dg_n e^{-i\theta_n}, \nn
& \theta_1 = -(\alpha_1 + \beta_1)/2, \nn
& \theta_2 = (\alpha_1 - \beta_1)/2, \nn
& \theta_{n+1} = \theta_n + \alpha_n \quad (n \geq 1), \nn
\end{align}
such that we now obtain the most general form of the Kitaev chain Hamiltonian
\begin{align}
H_{K} = \sum^N_{n=1} \varepsilon_n \widetilde{f}\dg_n \widetilde{f}_n + \sum^{N-1}_{n=1} (t_n \widetilde{f}\dg_{n+1} \widetilde{f}_n + \Delta_n e^{i\phi_n} \widetilde{f}\dg_{n+1} \widetilde{f}\dg_n + h.c.),
\end{align}
where $\phi_1=0$ and 
\begin{align}
 \phi_n = \phi_{n-1} + ( \beta_{n} - \beta_{n-1} ) + ( \alpha_{n} + \alpha_{n-1} )
\end{align}
for $n \geq 2$.
This indicates that phase plays no role in a minimal two-site chain and becomes important only when $N \geq 3$.

\section{Finite Zeeman spin splitting in the Andreev bound states}\label{sec:finite_zeeman}
\begin{figure}
\centering
{\includegraphics[width = 0.6\linewidth]{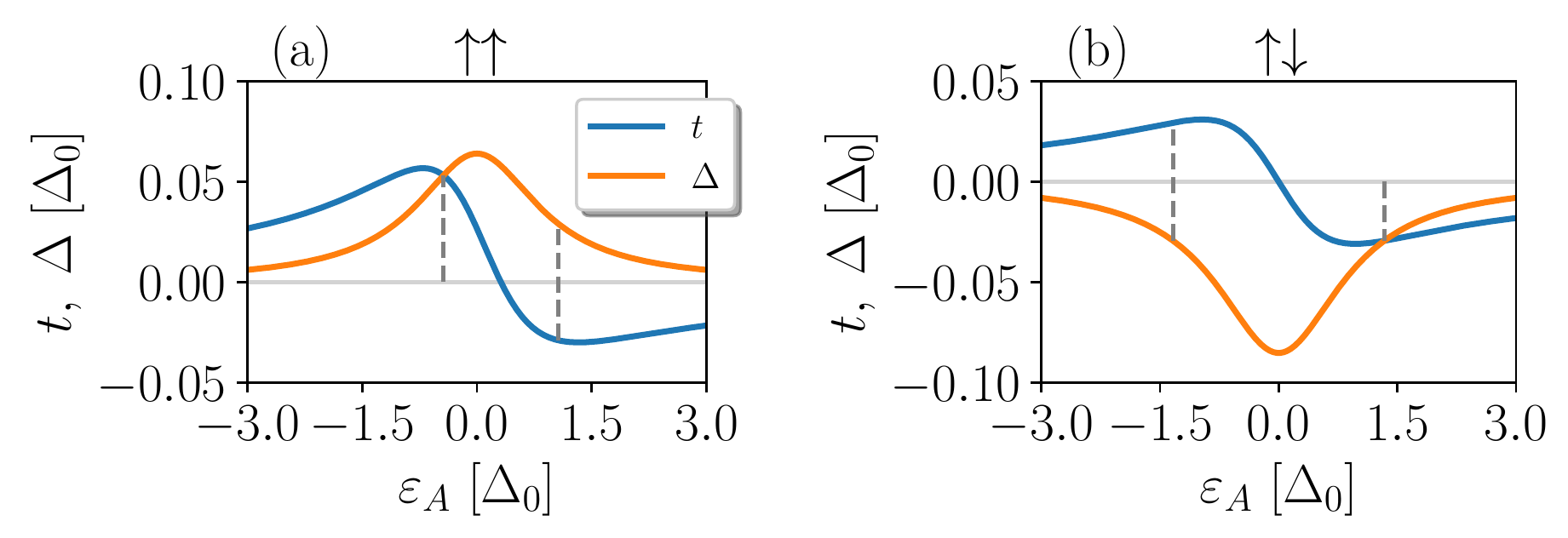}}
\caption{
CAR and ECT profiles in same- and opposite-spin channels in the presence of a weak Zeeman spin splitting ($E_Z/\Delta_0=0.25$) in ABS.
There still exist two sweet spots with opposite signs.
}
\label{fig:Fig_S1}
\end{figure}

In this section, we consider a weak Zeeman spin splitting in the Andreev bound state and show that the two opposite-sign sweet spots still exists. 
The only difference compared to zero-Zeeman scenario is that now the gap sizes at the two sweet spots become different.
The only modification we make in the Hamiltonian of Eq.~(3) is 
\begin{align}
H_{A,i} = \sum_{\sigma =\su, \sd} (\varepsilon_{Ai} + \sigma E_Z) n_{Ai\sigma} + ( \Delta_{i} c\dg_{Ai\su}c_{Ai\sd} + h.c.).
\end{align}
with $E_Z \ll \Delta$.
Using the second-order perturbation theory, the ECT and CAR between orbitals of same spins $H_{\text{eff}}(E_Z) = t_{\su \su} c\dg_{R\su} c_{L\su} + \Delta_{\su \su} c\dg_{R\su} c\dg_{L\su}+h.c.$ are 
\begin{align}
& t_{\su \su}(E_Z) = - t_{\rm{sc},1}  t'_{\rm{sc},1}  \left( \frac{u^2}{E_A + E_Z} - \frac{v^2}{E_A - E_Z}  \right)+ t_{\rm{sf},1}  t'_{\rm{sf},1} \left( \frac{u^2}{E_A - E_Z} - \frac{v^2}{E_A + E_Z}  \right) \nn
& \Delta_{\su \su}(E_Z) = ( t_{\rm{sc},1} t'_{\rm{sf},1} + t_{\rm{sf},1} t'_{\rm{sc},1} ) \left( \frac{uv}{E_A + E_Z} + \frac{uv}{E_A - E_Z}    \right).
\end{align}
Similarly, for the opposite-spin orbitals, we have
\begin{align}
&t_{\su \sd}(E_Z) = -  t_{1\rm{sc}} t'_{1\rm{sf}} \left(  \frac{u^2}{E_A + E_Z} - \frac{v^2}{E_A - E_Z}  \right)
- t_{1\rm{sf}} t'_{1\rm{sc}} \left(  \frac{u^2}{E_A - E_Z} - \frac{v^2}{E_A + E_Z}  \right), \nn
&\Delta_{\su \sd}(E_Z) =  ( t_{1\rm{sf}} t'_{1\rm{sf}}-  t_{1\rm{sc}} t'_{1\rm{sc}}  ) \left( \frac{uv}{E_A + E_Z} + \frac{uv}{E_A - E_Z}    \right).
\end{align}
The profiles of ECT and CAR for $E_Z/\Delta_0=0.25$ are shown in Fig.~\ref{fig:Fig_S1}.
For the same-spin channel, the two sweet spots are no longer symmetric about $\varepsilon_A$, they appear at $\varepsilon^*_A \approx -0.437\Delta_0$ and $\varepsilon^*_A \approx 1.062\Delta_0$.
For the opposite-spin channel, we still have $\varepsilon^*_A=\pm 4\Delta_0/3$.
We thus conclude that the findings of two sweet spots with opposite signs presented in the main text are robust against a weak Zeeman spin splitting in the ABS.
The presence of such a weak Zeeman field only modifies some details of the sweet spot, e.g., sweet-spot values of $\varepsilon_A$ and strength of $|t|$.

\begin{figure}[b]
    \centering
    \includegraphics[width = 0.6\linewidth]{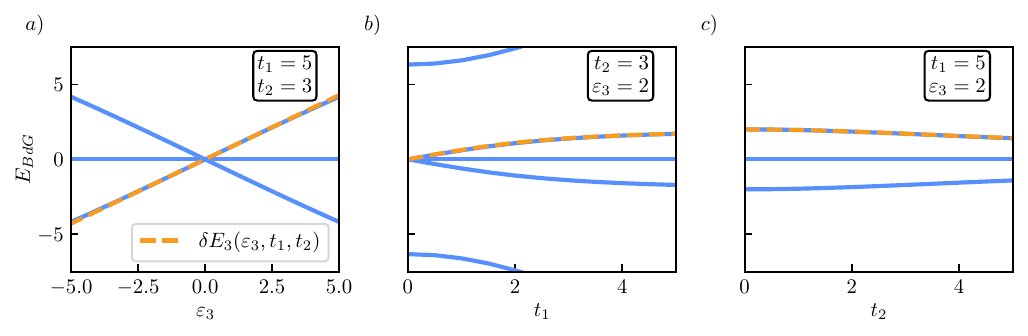}
    \caption{Energy spectra of Bogoliubov-de Gennes Hamiltonian in the vicinity of the sweet spot with $\phi=\pi$ as a function of $\varepsilon_3, t_1, t_2$.
    The numerically calculated splitting of the zero energy (solid blue lines) is in excellent agreement with the analytic results (dashed orange lines). }
    \label{fig:delta_e_numeric}
\end{figure}

\section{Splitting of zero modes at $\pi$ phase}
Here we provide a theoretical understanding for why the zero energy at $\pi$ phase gets split against detuning of the outermost dot.
Without loss of generality, we consider the spinless model for a three-site Kitaev chain.
The Bogoliubov-de Gennes Hamiltonian is
\begin{align}
H_3 =\bpm
\varepsilon_1 & t_1 & 0 & 0 & \Delta_1 & 0 \\
t_1 & \varepsilon_2 & t_2 & -\Delta_1 & 0 & \Delta_2 \\
0 & t_2 & \varepsilon_3 & 0 & -\Delta_2 & 0 \\
0 & -\Delta_1 & 0 & -\varepsilon_1 & -t_1 & 0 \\
\Delta_1 & 0 & -\Delta_2 & -t_1 & -\varepsilon_2 & -t_2 \\
0 & \Delta_2 & 0 & 0 & -t_2 & -\varepsilon_3
\epm 
=\bpm
0 & t_1 & 0 & 0 & t_1 & 0 \\
t_1 & 0 & t_2 & -t_1 & 0 & -t_2 \\
0 & t_2 & 0 & 0 & t_2 & 0 \\
0 & -t_1 & 0 & 0 & -t_1 & 0 \\
t_1 & 0 & t_2 & -t_1 & 0 & -t_2 \\
0 & -t_2 & 0 & 0 & -t_2 & 0 
\epm
\end{align}
under the basis of $(f_1,f_2,f_3,f_1^\dagger,f_2^\dagger,f_3^\dagger)$.
Here we choose $\varepsilon_1=\varepsilon_2=\varepsilon_3=0, t_1=\Delta_1$, and $t_2=-\Delta_2$ at $\pi$ phase.
There exist four Majorana zero modes at this point, with their wavefunctions being
\begin{align}
& \psi_1 = (1,0,0,1,0,0)^T/\sqrt{2}, \nn
& \psi_2 = (0,-i,0,0,i,0)^T/\sqrt{2}, \nn
& \psi_3 = (0,0,1,0,0,1)^T/\sqrt{2}, \nn
& \psi_4 = (it_2, 0, -it_1, -it_2, 0, it_1)^T/\sqrt{2(t^2_1+t^2_2)}.
\end{align}
Note that the first three Majoranas are localized at dots-1,2,3, while the fourth one is delocalized at the dots-1 and -3.
When the energy level of the third dot is detuned from resonance, the perturbation Hamiltonian is $H'=\text{diag}(0,0,\delta \varepsilon_3,0,0,-\delta \varepsilon_3)$, coupling the third and fourth Majorana modes.
The resulting energy splitting is therefore
\begin{align}
\delta E \approx | \langle \psi_3 | H' | \psi_4 \rangle| = \frac{t_1 }{ \sqrt{t^2_1+t^2_2} } \times \delta \varepsilon_3.
\end{align}
That is, to the leading order, the energy is split linearly, with the coefficient determined by the parameters $t_1$ and $t_2$.
To verify our analytic results, we numerically calculate the energy spectra of a three-site Kitaev chain model as a function of $\varepsilon_3, t_1, t_2$, as shown in Fig.~\ref{fig:delta_e_numeric}.
The numerical results are in excellent agreement with the analytic ones.

\end{document}